\documentclass[12pt]{article}
\usepackage{graphics}
\usepackage{amssymb}
\usepackage{amsmath}

\newcommand{\rf}[1]{(\ref{#1})}
\newcommand{\beq}{\begin{equation}}
\newcommand{\eeq}{\end{equation}}
\newcommand{\bea}{\begin{eqnarray}}
\newcommand{\eea}{\end{eqnarray}}

\begin{document}

\begin{center}

{ \Large \bf The Self-Organized de Sitter Universe}\\ 

\vspace{30pt}

{\sl J.\ Ambj\o rn}$\,^{a,c}$, 
{\sl J.\ Jurkiewicz}$\,^{b}$,
{\sl R.\ Loll}$\,^{c}$

\vspace{24pt}

{\footnotesize

$^a$~The Niels Bohr Institute, Copenhagen University\\
Blegdamsvej 17, DK-2100 Copenhagen \O , Denmark.\\
{ email: ambjorn@nbi.dk}\\

\vspace{10pt}

$^b$~Institute of Physics, Jagellonian University,\\
Reymonta 4, PL 30-059 Krakow, Poland.\\
{ email: jurkiewicz@th.if.uj.edu.pl}\\

\vspace{10pt}

$^c$~Institute for Theoretical Physics, Utrecht University, \\
Leuvenlaan 4, NL-3584 CE Utrecht, The Netherlands.\\
{ email: loll@phys.uu.nl}\\

}
\vspace{48pt}

{\sl 31 Mar 2008}

\vspace{20pt}

\end{center}


\begin{center}
{\bf Abstract}
\end{center}
We propose a theory of quantum gravity which formulates the quantum theory 
as a nonperturbative path integral, where each spacetime history 
appears with the weight $\exp{(iS^{\rm EH})}$, with $S^{\rm EH}$ the Einstein-Hilbert 
action of the corresponding causal geometry. The path integral is
diffeomorphism-invariant (only geometries appear) and background-independent. 
The theory can be investigated by computer simulations, which show that 
a {\it de Sitter universe} emerges on large scales. This emergence
is of an entropic, self-organizing nature, with the weight of the Einstein-Hilbert 
action playing a minor role. Also the quantum fluctuations around this 
de Sitter universe can be studied quantitatively and remain small 
until one gets close to the Planck scale. The structures found to describe 
Planck-scale gravity are reminiscent of certain aspects of condensed-matter systems.
\vspace{12pt}
\noindent


\newpage

\noindent{\large \bf De Sitter goes Quantum}\\ 

\noindent Little did Dutch astronomer Willem de Sitter realize in 1917 that his just discovered cosmological solution to Einstein's field equations \cite{desitter} would one day become an integral part of our description of the real, existing universe. What we today call the ``de Sitter universe" lay at the heart of de Sitter's famous debate with Einstein about the nature of spacetime at large, and the role played by boundaries and singularities \cite{roehle}. Neither man understood at the time what we nowadays think of as the defining feature of de Sitter space, namely, its dynamical nature. As is made explicit by the form
\beq\label{dsmetric}
ds^2=- d\tau^2+ \frac{3}{\Lambda}\cosh^2 
\left( \sqrt{\frac{\Lambda}{3}  }\ \tau \right) d\Omega^2_{(3)},
\eeq  
of the de Sitter metric ($d\Omega^2_{(3)}$ denotes the metric of the unit three-sphere), the distance between any two points will grow exponentially as proper time 
$\tau>0$ advances, with the expansion rate determined by the size of the cosmological constant 
$\Lambda$. After the discovery of the accelerated expansion of our universe \cite{expansion}, and the resurrection of the cosmological constant as likely dark energy source driving the expansion, we believe that the vacuum solution \rf{dsmetric} describes its inescapable fate in the far future, with all stars and galaxies apart from our own local galaxy cluster gradually fading from view \cite{krauss}. Besides providing a description of the universe at late times, de Sitter space also figures as a simplified model of the very early universe, as it undergoes rapid inflation after the big bang.

If all these developments were difficult to foresee almost 100 years ago, it would have been plain impossible -- before the advent of quantum mechanics -- to anticipate that de Sitter's universe would one day be reconstructed from nothing but quantum fluctuations! 
Here is what we can do in 2008 \cite{agjl}: a quantum ensemble of essentially structureless, microscopic constituents, which interact according to simple local rules dictated by gravity, causality and quantum theory, can produce a ``quantum universe", which on large scales matches perfectly É a classical, four-dimensional de Sitter universe! The derivation of this unprecedented result, obtained in the context of a candidate theory for quantum gravity based on Causal Dynamical Triangulations 
\cite{cdtreview}, is remarkable for a number of reasons:
\begin{itemize}
\item
It is genuinely background-independent: no preferred classical background metric is put into the construction at any stage.
\item It is genuinely nonperturbative: the path integral, a.k.a. the ``sum over histories" 
\beq\label{pi}
Z(G_N,\Lambda)=\int\limits_{\rm spacetime\atop geometries\ g\in\cal G}{\cal D}g\ {\rm e}^{iS^{\rm EH}[g]},
\;\;\;
S^{\rm EH}=\frac{1}{G_N}\int d^4x \sqrt{\det g} (R-2 \Lambda),
\eeq  
is dominated by spacetimes which are highly singular and nonclassical on short scales \cite{ajl}.
\item It is minimalist: no new fundamental objects (strings, loops, membranes, ...) or symmetry principles need to be postulated.  
\item It comes with a ``reality check": the quantum superposition \rf{pi} is not merely a formal quantity, but can be evaluated explicitly with the help of Monte Carlo simulations.
\item It is robust: many of the details of the intermediate regularization needed to make 
\rf{pi} mathematically well defined do not affect the final result, obtained in the continuum limit. 
\end{itemize}

\vspace{20pt}

\noindent{\large \bf Putting a New Spin on Quantum Gravity}\\ 

\noindent By general acknowledgement, deriving a classical limit from a full-fledged nonperturbative model of quantum gravity is exceedingly difficult and a major challenge for most approaches \cite{class}. How then does our method succeed in producing a ground state of quantum geometry, which is not only a classical spacetime on large scales, but -- in the absence of matter -- a physically realistic solution of the Einstein equations? Can the underlying theory answer longstanding questions about quantum gravity such as ``What are the true degrees of freedom of spacetime at the Planck scale?", and ``Can a smooth, classical spacetime ever emerge from microscopic, wild quantum fluctuations?"

Collecting all the evidence and results so far \cite{ajl,semi,agjl}, we believe our new formulation of quantum gravity yields important insights into how to think about gravity in the regime of ultra-short distances, usually captured by the heuristic notion of a ``spacetime foam" \cite{foam}. A fruitful approach is that of viewing quantum gravity {\it through the eyes of a condensed-matter theorist}, while paying close attention to key features of classical general relativity, like the need for coordinate-invariance and a causal structure. Think of quantum gravity as a strongly coupled system of a very large number of microscopic constituents, which by its nature is largely inaccessible to analytic pen-and-paper methods. This is no reason for dispair, but a common situation in many complex systems of theoretical interest in physics, biology and elsewhere, and merely calls for a dedicated set of technical tools and conceptual notions. 

The good news is that a relevant toolbox is at hand and contains powerful computational methods, which enable us to derive {\it quantitative} results. Their application relies on an intermediate discretization of the space of spacetime geometries, in the spirit of lattice spin systems or lattice QCD, but one that is coordinate-free and uses dynamical instead of fixed background lattices. A key aspect of the construction is that {\it if} a well-defined continuum limit of the path integral \rf{pi} exists as the discretization cut-off (or lattice mesh) is sent to zero (a nontrivial property that has to be demonstrated!), it will result in a fundamental theory valid on {\it all} scales.\footnote{Since in practice our computing power is always finite, we must take care that the finite lattice cut-off is much smaller than the smallest physical scale under study.}

\vspace{20pt}

\noindent{\large \bf The DIY Quantum Universe}\\ 

\noindent In concrete terms, what do you need to know to carry out the quantization program yourself, create a quantum universe on your laptop, and study its physical properties? There are straightforward construction rules for the spacetimes contributing to the regularized version of the path integral \rf{pi} \cite{ajl4d}: represent them as inequivalent piecewise flat manifolds (``triangulations") with a global proper-time structure, glued from four-dimensional triangular building blocks in a way that avoids causal singularities, like those associated with topology change. Next, set up a Monte Carlo simulation based on a Wick-rotated version of the path integral \rf{pi} and measure interesting quantum observables.   

How do you verify that the quantum superposition created by the computer behaves like a de Sitter universe? After having convinced yourself that it indeed behaves like a {\it four}-dimensional entity on large scales (something that cannot be taken for granted in nonperturbative quantum gravity \cite{bielefeld,semi}!), measure the expectation value $\langle N_3(t)\rangle$ of its spatial volume as a function of time $t$. The startling result is that one finds a universal curve $\langle N_3(t)\rangle \propto \cos^3 (const.\  t/N^{1/4})$, 
independent of the spacetime volume $N$. Translating this into a continuum language, and fixing one undetermined constant, the ratio between the time $t$ coming from the discrete triangulation and the ``true" proper time $\tau$ of the continuum formulation, this is seen to fit the shape of the de Sitter spacetime \rf{dsmetric} almost perfectly \cite{agjl}, see Fig.\ 1.
Note that we have substituted $\tau\mapsto i\tau$ in \rf{dsmetric} 
(which gives Euclidean de Sitter space, a four-sphere) in order to compare with
the computer simulations, which must be performed for the Wick-rotated, Euclideanized path integral.
\begin{figure}
\centerline{\scalebox{0.8}{\rotatebox{0}{\includegraphics{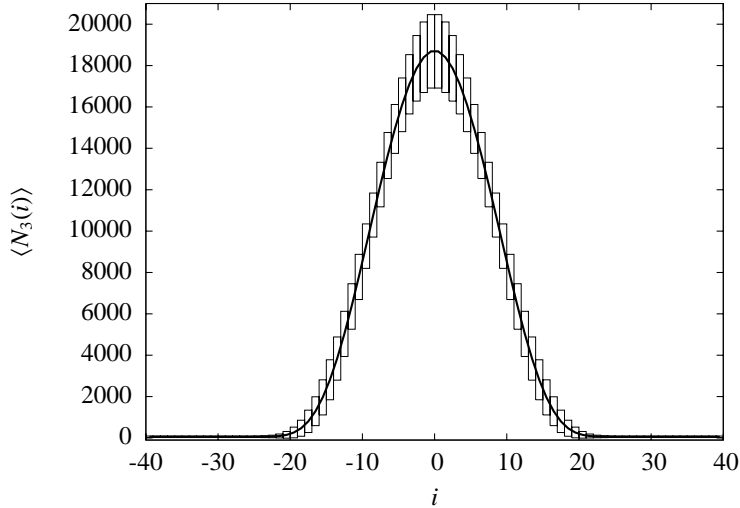}}}}
\caption{\label{fig1} \small Spatial volume $\langle N_3(i)\rangle$ as function of discrete time $i$: 
Monte Carlo measurements for fixed spacetime volume $N=362.000$
and best fit to the Euclideanized metric \rf{dsmetric} yield indistinguishable curves at this
plot resolution. 
The bars indicate the average size of quantum fluctuations, which also follow closely a
semi-classical calculation \cite{agjl}.}
\end{figure}

\vspace{30pt}

\noindent{\large \bf Complexity versus Simplicity}\\ 

\noindent Trying to understand the mechanism behind this miraculous ``emergence" of a (semi-)classical solution from the quantum theory, one quickly realizes that it is highly complex, as can be illustrated by comparing the relevant (Euclidean) actions. The ``bare" action of the path integral \rf{pi} is a straightforward discretization of the Einstein-Hilbert action \cite{regge} and shares its unboundedness from below\footnote{This is the infamous conformal-factor divergence \cite{conform}.}. It therefore induces a minisuperspace action 
\beq\label{mini}
S=\frac{1}{G_N}\int dt \left(-a(t) \dot a^2(t) -a(t) +\Lambda a^3(t) \right)
\eeq  
for the global scale factor $a(t)\sim N_3^{1/3}(t)$, with a characteristic kinetic term $\propto -\dot a(t)^2$, which has de Sitter space as a saddle point solution. By contrast, the effective action for $a(t)$ in the full quantum theory, arising from the {\it nonperturbative interplay between action and measure}, has a kinetic term $\propto +\dot a(t)^2$, and consequently the de Sitter solution there is a true minimum. 
Surprisingly, the collective {\it entropic effect} of all the gravitating degrees of freedom integrated over in the full path integral \rf{pi} leads effectively to an overall sign swap in the minisuperspace action \rf{mini} 
\cite{semi,ajl}! 
 
Despite the fact that our basic building blocks and interaction rules are simple, it is quite impossible to determine their combined dynamics analytically. In absence of any experiments probing Planck-scale geometry directly, to investigate the physical properties of such a strongly coupled quantum-gravitational system the use of numerical methods is therefore absolutely essential and should become part of every self-respecting quantum gravitator's toolbox! 

Borrowing a terminology from statistical and complex systems, we are dealing with a typical case of ``self-organization", a process where a system of a large number of microscopic constituents with certain properties and mutual interactions exhibits a collective behaviour, which gives rise to a new, coherent structure on a macroscopic scale.\footnote{Self-organized criticality in quantum gravity was previously considered in \cite{lee}.} What is particularly striking in our case is the recovery of a de Sitter universe, a {\it maximally symmetric space}, despite the fact that no symmetry assumptions were ever put into the path integral and we are employing a proper-time slicing \cite{ajl4d}, which na\"ively might have broken spacetime covariance. There clearly is much to be learned from this novel way of looking at quantum gravity!

\vspace{20pt}

\noindent{\bf Acknowledgment.}
We acknowledge support by
ENRAGE (European Network on
Random Geometry), a Marie Curie Research Training Network in the
European Community's Sixth Framework Programme, network contract
MRTN-CT-2004-005616. RL acknowledges
support by the Netherlands
Organisation for Scientific Research (NWO) under their VICI
program.
JJ acknowledges a partial support by the Polish Ministry of Science and
Information Technologies grant 1P03B04029 (2005-2008).

\end{document}